\documentclass[twocolumn,showpacs,preprintnumbers,amsmath,amssymb]{revtex4}
\usepackage{graphicx}
\usepackage{bm}
\usepackage{dcolumn}

\def\prl#1#2#3{{ Phys.   Rev.   Lett.  } {\bf #1}, #2 (#3)}

\def\pra#1#2#3{Phys.   Rev.   A {\bf #1}, #2 (#3)}
\def\svp#1#2#3{Sov. Phys.-JETP {\bf #1}, #2 (#3)}

\def\epl#1#2#3{{Europhys. Lett.} {\bf #1}, #2 (#3)}
\def\njp#1#2#3{{New Journal of Physics} {\bf #1}, #2 (#3)}
\def\pre#1#2#3{Phys.   Rev.   E {\bf #1}, #2 (#3)}
\def\pknaw#1#2#3{Prok. Kon. N. Akad. Wet {\bf #1}, #2 (#3)}
\def\ps#1#2#3{{Physica  } {\bf #1}, #2 (#3)}

\def\noi{\noindent}
\def\bc{\begin{center}}
\def\ec{\end{center}}
 \newcommand{\bea}{\begin{equation}}
 \newcommand{\eea}{\end{equation}\noi}
 \newcommand{\ber}{\begin{eqnarray}}
 \newcommand{\eer}{\end{eqnarray}\noi}
\begin{document}
\title{Bose-Einstein condensation and Casimir effect for ideal Bose Gas confined between two slabs}
\author{Shyamal Biswas}\email{tpsb@iacs.res.in}
\affiliation{Department of Theoretical Physics \\
Indian Association for the Cultivation of Science \\
Jadavpur,Kolkata-700032, India}

\date{\today}
\begin{abstract}
                We study the Casimir effect for a 3-d system of ideal Bose gas in a slab geometry with Dirichlet boundary condition. We calculate the temperature($T$) dependence of the Casimir force below and above the Bose-Einstein condensation temperature($T_c$). At $T\le T_c$ the Casimir force vanishes as $[\frac{T}{T_c}]^{3/2}$. For $T\gtrsim T_c$ it weakly depends on temperature. For $T\gg T_c$ it vanishes exponentially. At finite temperatures this force for thermalized photons in between two plates has a classical expression which is independent of $\hbar$. At finite temperatures the Casimir force for our system depends on $\hbar$.
\end{abstract}

\maketitle
           Vacuum fluctuation of electromagnetic field would cause an attractive force between two closely spaced parallel conducting plates. This phenomenon is called Casimir effect and this force is called Casimir force \cite{1,2,3}. In the original paper\cite{1} the Casimir force at zero temperature ($T=0$) was defined as 
\bea
F_c(L)=-\frac{\partial}{\partial L}[E(L)-E(\infty)]
\eea
where $E(L)$ is the ground state energy (i.e. the vacuum energy) of the electromagnetic field in between the two conducting plates separated at a distance $L$. This force has been measured experimentally \cite{4}. However, Casimir effect can be generalized \cite{5} for any range of temperature and for any dielectric substance  between two dielectric plates. It has also been generalized for thermodynamical systems \cite{6}. Casimir force for this kind of systems has recently been measured \cite{7}. At finite temperature $T$, the definition of Casimir force is generalized as \cite{8,9,10}
\bea
F_c(T,L)=-\frac{\partial}{\partial L}[\Omega_T(L)-\Omega_T(\infty)] 
\eea
where $\Omega_T(L)$ is the grand potential of the system confined between two plates separated at a distance $L$.

               We consider the Casimir effect for a thermodynamical system of Bose gas between two infinite slabs. Geometry of the system on which some external boundary condition can be imposed is responsible for the Casimir effect. Thermalized photons (massless bosons) in between two conducting plates of area $A$ at temperature $T$ gives rise to the Casimir pressure  \cite{11,12,13,14,15}
\begin{eqnarray}
\frac{F_c(L)}{A}\sim &-&\frac{\pi^{2}\hbar c}{240L^{4}}[1+\frac{16(kT)^{4}L^{4}}{3(\hbar c)^{4}}]\   \ for\  \frac{\pi\hbar c}{kTL}\gg 1 \nonumber\\ \sim &-&\frac{kT\zeta(3)}{8\pi L^{3}}\    \ for\  \frac{\pi\hbar c}{kTL}\rightarrow 0
\end{eqnarray}
where $k$ is the Boltzmann constant, $c$ is the velocity of light and $L$ is the separation of the parallel plates. At $T\rightarrow 0$, Casimir pressure becomes $-\frac{\pi^{2}\hbar c}{240L^{4}}$ and it is only the vacuum fluctuation which contributes to the Casimir pressure. At high temperature i.e. for $\frac{\pi\hbar c}{kTL}\rightarrow 0$, the Casimir force for photon gas  goes as $L^{-3}$ and has a purely classical expression independent of $\hbar$.

              Let us consider a Bose-gas is confined  between two infinitely large square shaped hard plates of area $A$. The plates are along x-y plane and they are separated along z- axis by a distance $L$. For the slab geometry, $\sqrt{A}\gg L$. We consider that our system is in thermodynamic equilibrium with its surroundings at temperature $T$. At this temperature the thermal de Broglie wavelength of a single particle of mass $m$ is $\lambda=\sqrt{\frac{\pi\hbar^{2}}{2mkT}}$. In the thermodynamic limit, $\frac{\lambda}{L}\ll 1$. For this system the single particle energy is $\epsilon(p_x,p_y,j)=\frac{p_x^2}{2m}+\frac{p_y^2}{2m}+\frac{\pi^2\hbar^2 j^2}{2mL^2}$, where $p_x$ and $p_y$ are the momentum along x-axis and y-axix respectively and $j=1,2,3,....$. However in the thermodynamic limit the single particle energy can be written as $\epsilon(p_x,p_y,p_z)=\frac{p_x^2}{2m}+\frac{p_y^2}{2m}+\frac{p_z^2}{2m}$, where $p_z$ is the momentum along z-axis.

             Considering the thermodynamic limit the total number of thermally excited particles can be written as
\bea
N_T=\int_{-\infty}^{\infty}\int_{-\infty}^{\infty}\int_{-\infty}^{\infty}\frac{1}{e^{\frac{[\frac{p_x^{2}}{2m}+\frac{p_x^{2}}{2m}+\frac{p_z^{2}}{2m}-\mu]}{kT}}-1}\frac{Vdp_xdp_ydp_z}{[2\pi\hbar]^{3}}
\eea
where $\mu$ is the chemical potential and $V$ is the volume of the system. Bose condensation temperature ($T_c$) is defined as a temperature where all the particles are thermally excited and below that temperature a macroscopic number of particles come to the ground state \cite{16,17,18}. At $T\le T_c$ the chemical potential goes to the ground state energy. So
\begin{eqnarray}
N&=&\int_{-\infty}^{\infty}\int_{-\infty}^{\infty}\int_{-\infty}^{\infty}\frac{1}{e^{\frac{[\frac{p_x^{2}}{2m}+\frac{p_x^{2}}{2m}+\frac{p_z^{2}}{2m}]}{kT}}-1}\frac{Vdp_xdp_ydp_z}{[2\pi\hbar]^{3}}\nonumber\\ &=&\frac{V}{[2\pi\hbar]^{3}}(2\pi mkT)^{3/2}\sum_{i=1}^{\infty}\frac{1}{i^{3/2}}\nonumber\\&=&\frac{V}{8}\frac{1}{\lambda_c^3}\zeta(3/2)\nonumber\\&=&\frac{AkT_cm}{2\pi\hbar^2}\frac{L}{2\lambda_c}\zeta(3)
\end{eqnarray}
where $\lambda_c =\sqrt{\frac{\pi\hbar^{2}}{2mkT_c}}$. Now from equation(5) we have
\bea
T_c=\frac{1}{k}[\frac{2\pi\hbar^{2}}{m}][\frac{N}{V\zeta(3/2)}]^{\frac{2}{3}}
\eea

                Let us now introduce the finite size correction. The ground state energy of our system is $[g=\frac{\pi^{2} \hbar^{2}}{2mL^{2}}]$. The average no. of particles with energy $\epsilon_{p_x,p_y,j}$ is given by $\frac{1}{e^{[\frac{p_x^2}{2m} +\frac{p_y^2}{2m}+\frac{\pi^{2} \hbar^{2}(j^{2}-1)}{2mL^{2}}-\mu']/kT}-1}$ where $\mu'=(\mu-g)\le 0$ for bosons. At and below the condensate temperature $\mu'\rightarrow 0$. For this bosonic system we have the grand potential 
\begin{eqnarray}
&&\Omega=\Omega(A,L,T,\mu')\nonumber\\
&&=kT\sum_{j=1}^{\infty}\int_{-\infty}^{\infty}\int_{-\infty}^{\infty}[\frac{Adp_xdp_y}{[2\pi\hbar]^2}\nonumber\\&&\log [1-e^{\frac{-(\frac{p_x^2}{2m} +\frac{p_y^2}{2m}+\frac{\pi^{2}\hbar^{2}(j^{2}-1)}{2m L^{2}}-\mu')}{kT}}]]
\end{eqnarray}
Replacing $j$ by $(j'+1)$, we recast the above equation as
\begin{eqnarray}
&&\Omega(\omega,L,T,\mu')\nonumber\\&&=-kT\sum_{i=1}^{\infty}\int_{p_x=0}^{\infty}\int_{p_y=0}^{\infty}\sum_{j'=0}^{\infty}\frac{Adp_xdp_y}{[2\pi\hbar]^2}\nonumber\\&&\frac{e^{\frac{i\mu'}{kT}}e^{-\frac{ip_x^2}{2mkT}}e^{-\frac{ip_y^2}{2mkT}}e^{-i(\pi(\frac{\lambda}{L})^{2}[j'^{2}+2j'])}}{i}
\end{eqnarray}
where $\lambda =\sqrt{\frac{\pi\hbar^{2}}{2mkT}}$.
Integrating over $p_x$ and $p_y$ we get
\begin{eqnarray}
&&\Omega(A,L,T,\mu')\nonumber\\&&=-\frac{AkT}{(2\pi\hbar)^2}[2\pi mkT]\sum_{i=1}^{\infty}\sum_{j'=0}^{\infty}\frac{e^{i\mu'/kT}}{i^{2}}[e^{\frac{-\pi i \lambda^{2}(j'^{2}+2j')}{L^{2}}}]
\nonumber\\&&=-\frac{A(kT)^2 m}{2\pi\hbar^2}\sum_{i=1}^{\infty}\sum_{j'=0}^{\infty}\frac{e^{i\mu'/kT}}{i^{2}}[e^{-\frac{\pi i \lambda^{2}j'^{2}}{L^{2}}}][1\nonumber\\&&-2j'\frac{\pi i \lambda^{2}}{L^{2}}+2j'^{2}(\frac{\pi i \lambda^{2}}{L^{2}})^{2}-\frac{4}{3}j'^{3}(\frac{\pi i \lambda^{2}}{L^{2}})^{3}+...]
\end{eqnarray}
Since $\frac{\lambda}{L}\ll 1$, higher order terms of the above series would not contribute significantly. From Euler-Maclaurin summation formula we convert the summation over $j'$ to integration. So from equation(9) we have
\begin{eqnarray}
&&\Omega(A,L,T,\mu')\nonumber\\&&=-\frac{A(kT)^2m}{2\pi\hbar^2}\sum_{i=1}^{\infty}\frac{e^{i\mu'/kT}}{i^{2}}[(\int_{0}^{\infty}e^{\frac{-\pi i \lambda^{2}j'^{2}}{L^{2}}}dj'+\frac{1}{2})-\nonumber\\&&2\frac{\pi i \lambda^{2}}{L^{2}}(\int_{0}^{\infty}j'e^{-\frac{\pi i \lambda^{2}j'^{2}}{L^{2}}}dj'-\frac{1}{12})+2[\frac{\pi i \lambda^{2}}{L^{2}}]^{2}\nonumber\\&&(\int_{0}^{\infty}j'^{2}e^{-\frac{\pi i \lambda^{2}j'^{2}}{L^{2}}}dj')-\frac{4}{3}[\frac{\pi i \lambda^{2}}{L^{2}}]^{3}(\int_{0}^{\infty}j'^{3}e^{-\frac{\pi i \lambda^{2}j'^{2}}{L^{2}}}dj'\nonumber\\&&+\frac{6}{720})+...] 
\end{eqnarray}
Collecting the leading terms from the above equation(10) we can write
\begin{eqnarray}
&&\Omega(A,L,T,\mu')\nonumber\\&&=-\frac{A(kT)^2m}{2\pi\hbar^2}\sum_{i=1}^{\infty}\frac{e^{i\mu'/kT}}{i^{2}}[\frac{L}{2\lambda i^{1/2}}-\frac{1}{2}+\nonumber\\&&\frac{\pi}{2}i^{1/2}\frac{\lambda}{L}+\it{O}([\frac{\lambda}{L}]^{2})]\nonumber\\&&=-\frac{A(kT)^2m}{2\pi\hbar^2}[\frac{L}{2\lambda}g_{\frac{5}{2}}(z)-\frac{1}{2}g_{2}(z)+\frac{\pi\lambda}{2L}g_{\frac{3}{2}}(z)]\nonumber\\
\end{eqnarray}
where $z=e^{\mu'/kT}$ is the fugacity and $g_l(z)=z+\frac{z^2}{2^{l}} +\frac{z^3}{3^{l}}+....$ is the Bose-Einstein function. From the above equation we get the total number of particles as 

\begin{eqnarray}   
&&N=-\frac{\partial\Omega}{\partial\mu'}\nonumber\\&&=\frac{AkTm}{2\pi\hbar^2}[\frac{L}{2\lambda}g_{\frac{3}{2}}(z)-\frac{1}{2}g_{1}(z)+\frac{\pi\lambda}{2L}g_{\frac{1}{2}}(z)]
\end{eqnarray}
In the thermodynamic limit of a system, as $T\le T_c$, $z\rightarrow 1$. For a finite system this can not happen, otherwise the correction terms in the above expression would be infinite. Instead at $T\gtrsim T_c$, $z\sim 1$. Taking only the first correction term in the eqn.(12) we have $g_1(z)=-ln(1-z)=[N'(T)g_{\frac{3}{2}}(z)-N\zeta(3/2)]\frac{L}{\lambda}=-ln\triangle z$, where $N'(T)= \frac{AkTm}{2\pi\hbar^2}[\frac{L}{2\lambda}\zeta(3/2)]$ and $\triangle z=1-z$ is a small change in the fugacity at $T\gtrsim T_c$. Now putting $z=1$ in the expression of $\triangle z$, we get $\triangle z=e^{-\triangle N\zeta(3/2)L/\lambda}$, where $\triangle N=N'(T)-N$. We see that in the thermodynamic limit($L\rightarrow\infty$) $\triangle z=0$ and when $L$ is finite such that $L/\lambda\gg 1$ we have $z\sim 1$ at $T\gtrsim T_c$.

               Let us now calculate the Casimir force. At $T\le T_c$ we put $\mu'\rightarrow 0$ or $z\rightarrow 1$. So from eqn.(11) we have
\begin{eqnarray}
\Omega(\omega,L,T,0)
 =&-&\frac{A(kT)^2m}{2\pi\hbar^2}[\frac{L}{2\lambda}\zeta(5/2)-\frac{1}{2}\zeta(2)\nonumber\\&&
+\frac{\pi}{2}\zeta(3/2)\frac{\lambda}{L}]
\end{eqnarray}

 Here the first term of eqn. (13) is
\bea
\Omega_b=-\frac{A(kT)^2m}{2\pi\hbar^2}[\frac{L}{2\lambda}\zeta(5/2)].
\eea
It is the bulk term of the grand potential. From our consideration of thermodynamic limit $\frac{AL}{N}=constant$. So $\Omega_T(\infty)=\Omega_b$. The second term of eqn. (13) is $(\Omega_s)=\frac{A(kT)^2m}{2\pi\hbar^2}[\frac{1}{2}\zeta(2)]$. It is the surface term of the grand potential. The third term of eqn.(13) is the Casimir term of the grand potential. We call it the Casimir potential. Now putting $N =\frac{AkT_cm}{2\pi\hbar^2}\frac{L}{2\lambda_c}\zeta(3/2)$ in eqn.(13) we find the Casimir potential as
\begin{eqnarray}
\Omega_c=&-&\frac{A(kT)^2m}{2\pi\hbar^2}\frac{L}{\lambda}\zeta(3/2)\pi(\frac{\lambda}{L})^2\nonumber\\=&-&NkT\pi(\frac{\lambda}{L})^2(\frac{T}{T_c})^{3/2}
\end{eqnarray}
Putting $\lambda=\sqrt{\frac{\pi\hbar^2}{2mkT}}$ in equation(15) we have
\bea
\Omega_c=-N[\frac{T}{T_c}]^{3/2}\frac{\pi^{2}\hbar^{2}}{2mL^{2}}
\eea
From eqn.(2) and (14) we have
\bea
F_c(T,L)=-\frac{\partial}{\partial L}\Omega_c
\eea
For $T\le T_c$, from eqn.(16) and (17) we have the expression of Casimir force as
\bea
F_c(T,L)=-N[\frac{T}{T_c}]^{3/2}\frac{\pi^{2}\hbar^{2}}{mL^{3}}
\eea
This expression for the Casimir force shows that, at finite temperatures the force depends on $\hbar$.

            Above the condensation temperature $\mu'<0$ or $z<1$. However, for $T\gtrsim T_c$, $z\sim 1$. So at $T\gtrsim T_c$, from the eqn.(11) and eqn.(5) with trivial manipulation we get the Casimir potential as 
\begin{eqnarray}
\Omega_c&=&-\frac{A(kT)^2m}{2\pi\hbar^2}\frac{\pi\lambda}{2L}g_{\frac{3}{2}}(z)\nonumber\\&\approx& -\pi N kT (\frac{\lambda}{L})^{2}(\frac{T}{T_c})^{3/2} 
\end{eqnarray}
In the above equation for $T\gtrsim T_c$, we put $z=1$. However as $\frac{T}{T_c}$ increases $z$ decreases. So, for $T\gtrsim T_c$ the Casimir potential weakly depends on temperature.
Putting $T=T_c+\triangle T$ in equation (19) and from the definition of Casimir force we have 
\bea
F_c(T,L)\approx-N\frac{\pi^{2}\hbar^{2}}{mL^{3}}
\eea 
where $0<\frac{\triangle T}{T_c}\ll 1$. For $T\gtrsim T_c$, the Casimir force weakly depends on temperature.

\begin{figure}
\includegraphics{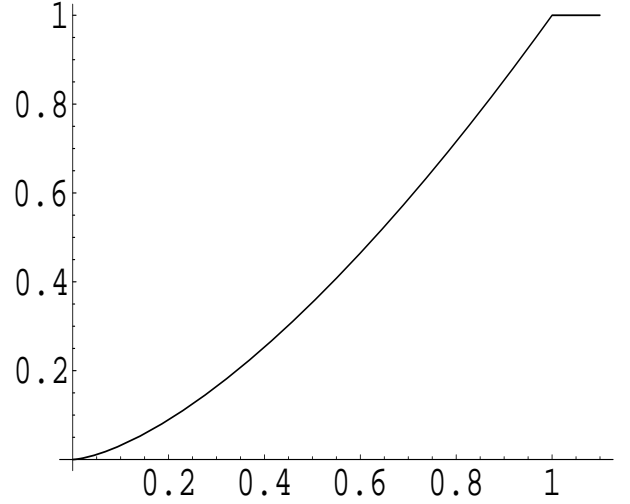}
\caption {Casimir force($F_c(T,L)$) versus temperature($T$) plot. $F_c(T,L)$ is in units of $N\frac{\pi^2\hbar^2}{mL^3}$ and $T$ is in units of $T_c$. Equation (18) corresponds to the range $\frac{T}{T_c}\le 1$. Equation (20) corresponds to the range $1<\frac{T}{T_c}\le 1.1$.} 
\label{fig:Casimir Force}
\end{figure} 

            Let us now calculate the Casimir force at $T\gg T_c$. At these temperatures $z\ll 1$. So we can approximately write $g_i(z)\approx z+\frac{z^2}{2^i}$. From the first term of eqn.(12) we have $g_{\frac{3}{2}}(z)=\frac{N2\pi\hbar^2 2\lambda}{AkTmL}\approx z+z/(2\sqrt{2})$. For this range of temperatures we can write $e^{\mu/kT}=ze^{-\pi\lambda^2/L^2}\approx z$. For convenience, we replace $\mu'$ by $\mu$ and recast equation (7) as
\begin{eqnarray}
&&\Omega=\Omega(A,L,T,\mu)\nonumber\\
&&=-kT\sum_{j=1}^{\infty}\int_{-\infty}^{\infty}\int_{-\infty}^{\infty}[\frac{Adp_xdp_y}{[2\pi\hbar]^2}\nonumber\\&&\log [1-e^{\frac{-(\frac{p_x^2}{2m} +\frac{p_y^2}{2m}+\frac{\pi^{2}\hbar^{2}j^{2}}{2m L^{2}}-\mu)}{kT}}]]\nonumber\\&&=-\frac{A(kT)^2m}{2\pi\hbar^2}\sum_{i=1}\sum_{j=1}\frac{e^{\mu i/kT}}{i^2}e^{-i\pi\lambda^2 j^2/L^2}\nonumber\\&&=-\frac{A(kT)^2m}{2\pi\hbar^2}\sum_{i=1}\sum_{j=1}\frac{e^{\mu i/kT}}{i^2}(\frac{L}{2\lambda i^{\frac{1}{2}}}-\frac{1}{2}+\frac{L}{\lambda i^{\frac{1}{2}}}e^\frac{-\pi L^2j^2}{i\lambda^2})\nonumber\\
\end{eqnarray}
where we use the formula $\sum_{n=-\infty}^{\infty}e^{-\pi an^2}=\frac{1}{\sqrt{a}}\sum_{n=-\infty}^{\infty}e^{-\pi n^2/a}$. From the above equation we choose the Casimir potential as $\Omega_c=-\frac{A(kT)^2m}{2\pi\hbar^2}\sum_{i=1}^{\infty}\sum_{j=1}^{\infty}\frac{e^{\mu i/kT}}{i^{5/2}}\frac{L}{\lambda}e^\frac{-\pi L^2j^2}{i\lambda^2}$. For $\frac{T}{T_c}\rightarrow\infty$, in the expression of above Casimir potential we can  put $e^{\mu/kT}=z\ll 1$ and can take $i=1$ and $j=1$ as the leading term to contribute in the Casimir potential. So, for $T\gg T_c$ the Casimir potential is
\begin{eqnarray}
\Omega_c&=&-\frac{A(kT)^2m}{2\pi\hbar^2}\frac{L}{\lambda}\frac{e^{\mu /kT}}{1^{5/2}}e^{-\pi L^2/\lambda^2}\nonumber\\&=&-2NkTe^{-\frac{2mL^2kT}{\hbar^2}}
\end{eqnarray}
where we put $e^{\mu/kT}=z\approx g_{\frac{3}{2}}(z)=\frac{N2\pi\hbar^2 2\lambda}{AkTmL}$.
From equation (22), for $T\gg T_c$ we have the Casimir force as 
\begin{eqnarray}
F_c(T,L)&=&-\frac{\partial\Omega_c}{\partial L}\nonumber\\&=&-\frac{8N(kT)^2mL}{\hbar^2}e^{-\frac{2mL^2kT}{\hbar^2}}
\end{eqnarray}
  
Now we see that in the classical limit($T\gg T_c$) the Casimir force vanishes as $e^{-kT}$. 

             The changes of Casimir force with temperature for the range $0<T\le T_c$ and for the range $T\gtrsim T_c$ is shown in FIG 1. 
        
             That vacuum fluctuation causes Casimir force is well known\cite{1,4}. Critical fluctuation also causes Casimir force\cite{7,9}. The Casimir force calculated here is neither due to vacuum fluctuation nor due to critical fluctuation. It is due to quantum fluctuation. This fluctuation is associated with the commutator algebra of position and momentum operator as well as with the commutator algebra of bosonic annihilation operator($\hat{a}_i$) and creation operator ($\hat{a}_i^{\dagger}$) such that $[\hat{a}_i,\hat{a}_j^{\dagger}]=\delta_{i,j}$, where $i,j$ represent the single particle energy states. At $T\ll T_c$ almost all the particles come down to the ground state. The quantum fluctuation dies out due to the macroscopic occupation of particles in a single state. That is why the Casimir force dies out at $T\ll T_c$. At $T\gg T_c$ the Bose-Einstein statistics becomes classical Maxwell-Boltzmann statistics and thermal fluctuation dominates over the quantum fluctuation. For this reason the Casimir force dies out at $T\gg T_c$. Below $T_c$ the reduction of thermodynamic Casimir force with $T^{3/2}$ law is the signature of Bose-Einstein condensation. 
          
           Several useful discussions with J. K. Bhattacharjee and Sudipto Paul Chowdhury of I.A.C.S. are gratefully acknowledged.

\end{document}